\documentclass[12pt]{article}
\usepackage{graphics}
\usepackage{latexsym, pstricks}
\usepackage{amsmath,amssymb,amsthm}

\font\titulo=cmbx10 scaled\magstep1 
scaled\magstep1

\def\section#1{\vskip 1.5truepc plus 0.1truepc minus 0.1truepc
    \goodbreak \leftline{\titulo#1} \nobreak \vskip 0.1truepc
    \indent}
\def\frc#1#2{\leavevmode\kern.1em
    \raise.5ex\hbox{\the\scriptfont0 $ #1 $}\kern-.1em
    /\kern-.15em\lower.25ex\hbox{\the\scriptfont0 $ #2 $}}

\def\IZ{{\rm Z}\llap{\vrule height7.1pt width1pt
     depth-.4pt\phantom t}} 

\newbox\pmbbox
 \def\pmb#1{{\setbox\pmbbox=\hbox{$#1$}%
\copy\pmbbox\kern-\wd\pmbbox\kern.3pt\raise.3pt\copy\pmbbox\kern-\wd\pmbbox
\kern.3pt\box\pmbbox}}

\font\cmss=cmss10 \font\cmsss=cmss10 at 7pt

\def\IZ{\relax\ifmmode\mathchoice
{\hbox{\cmss Z\kern-.4em Z}}{\hbox{\cmss Z\kern-.4em Z}}
{\lower.9pt\hbox{\cmsss Z\kern-.4em Z}} {\lower1.2pt\hbox{\cmsss
Z\kern-.4em Z}}\else{\cmss Z\kern-.4em Z}\fi}

\font\cmss=cmss10 \font\cmsss=cmss10 at 7pt
\def\IS{\relax\ifmmode\mathchoice
{\hbox{\cmss S\kern-.4em S}}{\hbox{\cmss S\kern-.4em S}}
{\lower.9pt\hbox{\cmsss S\kern-.4em S}} {\lower1.2pt\hbox{\cmsss
S\kern-.4em S}}\else{\cmss S\kern-.4em S}\fi}

\begin{document}

\centerline{\titulo Phase Transition in \textit{NK}-Kauffman Networks}

\centerline{\titulo and its Correction for Boolean Irreducibility}

\vskip 1.2pc \centerline{Federico Zertuche}

\vskip 1.2pc \centerline{Instituto de Matem\'aticas, Unidad
Cuernavaca} \centerline{Universidad Nacional Aut\'onoma de
M\'exico} \centerline{A.P. 273-3, 62251 Cuernavaca, Mor.,
M\'exico.} \centerline{\tt federico.zertuche@im.unam.mx}

\vskip 3pc {\bf \centerline {Abstract}}

In a series of articles published in 1986 Derrida, and his colleagues studied two mean field treatments (the quenched and the annealed) for \textit{NK}-Kauffman Networks. Their main results lead to a phase transition curve $ K_c \, 2 \, p_c \left( 1 - p_c \right) = 1 $ ($ 0 < p_c < 1 $) for the critical average connectivity $ K_c $ in terms of the bias $ p_c $ of extracting a ``$1$" for the output of the automata. Values of $ K $ bigger than $ K_c $ correspond to the so-called chaotic phase; while $ K < K_c $, to an ordered phase. In~[F. Zertuche, {\it On the robustness of NK-Kauffman networks against changes in their connections and Boolean functions}. J.~Math.~Phys. {\bf 50} (2009) 043513], a new classification for the Boolean functions, called {\it Boolean irreducibility} permitted the study of new phenomena of \textit{NK}-Kauffman Networks. In the present work we study, once again the mean field treatment for \textit{NK}-Kauffman Networks, correcting it for {\it Boolean irreducibility}. A shifted phase transition curve is found. In particular, for $ p_c = 1 / 2 $ the predicted value $ K_c = 2 $ by Derrida {\it et al.} changes to $ K_c = 2.62140224613 \dots $ We support our results with numerical simulations.

\vskip 2pc

\noindent {\bf Short title:} {\it Boolean Irreducibility and Phase Transitions}

\vskip 1pc \noindent {\bf Keywords:} {\it Cellular automata,
Boolean irreducibility, binary functions, phase transitions,
\textit{NK}-Kauffman Networks}.


\newpage

\baselineskip = 12.4pt

\section{1. Introduction}

\textit{NK}-Kauffman networks have been widely studied due to their applications in theoretical biology; they are specially useful in the study of the genotype-phenotype map $ \Psi $~[1-4]. As long as they are randomly constructed, they are well suited for study with statistical techniques; in particular by the use of mean field approximations.

A \textit{NK}-Kauffman network consists of $ N $ Boolean variables $ S_i (t) \in \mathbb{Z}_2 $ ($ i = 1, \dots, N $), which evolve deterministically in discrete time $ t = 0, 1, 2, \dots $ according to Boolean functions on $ K $ ($ 0 \leq K \leq N $) of these variables at the previous time $ t-1 $. For every site $ i $, a $K$-Boolean function $ \mathfrak{F}_i: \mathbb{Z}_2^K \to \mathbb{Z}_2 $ is randomly and independently chosen with a bias probability, $ p $:
$$
0 < p < 1 , \eqno(1)
$$
that $ \mathfrak{F}_i = 1 $; and $ \mathfrak{F}_i = 0 $ with probability $ 1 - p $ for each of its $ 2^K $ possible arguments. Also, for every site $ i $, $ K $ inputs $ i_1, \ldots, i_K $ (the connections) are randomly selected, without repetition, from a uniform distribution among the $ N $ Boolean variables of the network. So, for each site $ i $, and each extraction $ E $ ($ 0 \leq E \leq K - 1 $); an input $j$ ($ 1 \leq j \leq N $) is obtained with probability
$$
\texttt{P}_{i,E} \left( j \right) = \left\{ \begin{array}{ll}
{1 \over N - E} , & \hskip0.15cm \mbox{if $j$ has not been extracted}
\\ {} & {} \\ 0 , & \hskip0.15cm \mbox{otherwise} \end{array} \right. \eqno(2)
$$

Once the $ K $ inputs and the functions $ \mathfrak{F}_i $ have been selected, a Boolean deterministic \textit{NK}-Kauffman network has been defined. So, what we have obtained is a dynamical system that evolves deterministically, and synchronously in time, according to the rules
$$
S_i (t+1) = {\mathfrak F}_i \left( S_{i_1}(t), S_{i_2}(t), \dots, S_{i_K}(t)
\right), \ \ i = 1, \dots, N, \eqno(3)
$$
where $ i_m \not= i_n $, for all $ m, n = 1, 2, \dots, K $, and $ m \not= n $ since from (2), each input is different. \textit{NK}-Kauffman networks are a special type of Boolean endomorphism $ f: \mathbb{Z}_2^N \to \mathbb{Z}_2^N $. Let $ {\cal B}_N $, denote the set of Boolean endomorphisms, and $ {\cal L}^N_K $ the set of \textit{NK}-Kauffman networks. Then $ {\cal L}^N_K \subseteq {\cal B}_N $, and $ {\cal L}^N_N \cong {\cal B}_N $~[3,4]. Note also that $ {\cal B}_N \cong {\cal G}_{2^N} $: where $ {\cal G}_{2^N} $ is the set of functional graphs with $ 2^N $ points {\it i.e.} the directed graphs with {\tt out-degree one}, and loops allowed~[5]. In Ref.~[3], a study of the injective properties of the map
$$
\Psi : {\cal L}^N_K \rightarrow {\cal B}_N \cong {\cal G}_{2^N} \eqno(4)
$$
was done, and enabled the calculation of the average number $ \vartheta \left( N, K \right) $ of elements in $ {\cal L}^N_K $ that $ \Psi $ maps into the same functional graph~[3,4]. The results showed that there exists a critical average connectivity $ \hat{K} $ for $ N \gg 1 $, given by
$$
\hat{K} \approx \log_2 \log_2 \left( {2 N \over \ln 2} \right) + {\cal
O} \left( {1 \over N \ln N } \right); \eqno(5)
$$
such that $ \vartheta \left( N, K \right) \approx e^{\varphi \, N} \gg 1 $ ($ \varphi > 0 $) or $ \vartheta \left( N,K \right) \approx 1 $, depending on whether $ K < \hat{K} $ or $ K > \hat{K} $, respectively. That is to say, $ \Psi $ is almost an injective function for $ K > \hat{K} $, and almost a many-to-one function for $ K < \hat{K} $.

An important challenge, since the proposal by Kauffman about his networks~[1,2], has been the analytic calculation of their average dynamics in terms of the network parameters, which are: the number $ N $ of Boolean variables, their connectivity $ K $, and the extraction bias $ p $ of the Boolean functions $ {\mathfrak F}_i $. Until now, only some special cases have been analytically solved. Among the most important are:

The extreme equiprobable cases ($ p = 1/2 $), with  $ K = N $, the so-called {\it random map model}~[5,6], and $ K = 1 $~[7]. And, of particular interest, the case of $ K = 2 $, with $ p \neq 1 / 2 $, and $ p = 1/2 $; which was studied by the use of random graphs techniques, and combinatorial methods in a series of articles by Lynch, with remarkable results~[8].

In 1986, in a series of works Derrida {\it et al.} studied extensively Kauffman's model, and some of its variations~[9-11]: Derrida \& Pomeau studied an {\it annealed approximation} for the case $ p = 1/2 $~[9]. Such model differs from Kauffman's model ({\it the quenched case}) in that the connections and Boolean functions $ \mathfrak{F}_i $ are shuffled at each time step. While the annealed model exhibits a different dynamical behavior in relation to Kauffman's (for example, limit cycles are absent in it) both models exhibit phase transitions at the same value $ K_c = 2 $ for $ p = 1/2 $, a fact that was well supported with numerical simulations~[9]. Furthermore, while Ref.~[11] is mainly devoted to a mean field approximation of a cellular automata model in a two dimensional lattice, in that article, is also shown, how, for the case of Kauffman's model (the infinite dimensional lattice case in that work) the phase transition equation may be generalized for the biased case (1). The conclusions are well summarized by Derrida in Ref.~[12] showing that there is a critical connectivity $ K_c $, such that Hamming distance between two nearby states grows or decays exponentially according to whether $ K > K_c $ or $ K < K_c $, respectively; with the critical curve given by~[11-13]
$$
K_c \ 2 \, p_c \left( 1 - p_c \right) = 1 . \eqno(6)
$$

In Ref.~[4], a classification of Boolean functions according to the number of arguments that really influence on the functions' output was proposed. It was called the {\it irreducible degree classification} of Boolean functions. By its means several features of \textit{NK}-Kauffman networks have been calculated: In Refs.~[3,4] the critical transition (5) was calculated . In Ref.~[4] the robustness of \textit{NK}-Kauffman networks (3) against random changes of $ {\mathfrak F}_i $ and its connections $ i_1, \ldots, i_K $ was performed. The results were in good agreement with experimental studies of the genetic material by induced mutations for the all important case $ K \simeq 2 $ for specific values of $ p $. In Ref.~[14] algebraic techniques were used to calculate the number of $K$-Boolean functions with a $ \lambda $-degree of irreducibility and weight $ \omega $, denoted by $ \varrho_K \left( \lambda, \omega \right) $. This quantity plays a fundamental role for the calculations of the present work.

The scope of the present work is to make a mean field analysis of Kauffman's model: defined by Eqs.~(1), (2), and (3) taking into account the Boolean irreducibility. As we show, the transition curve obtained approaches asymptotically to the curve (6) for values of $ p $ in the extreme zones $ p \sim 0 $ and $ p \sim 1 $. On the contrary for the zone where $ p \sim 1/2 $ the irreducibility effect of the Boolean functions becomes more pronounced, and the corrected-for-irreducibility curve gets its maximum difference from (6) at $ p = 1/2 $.

Several simulations have been done for Kauffman's model for the $ p = 1/2 $ case~[1,15]. The main result was that the critical connectivity should be on the range $ 2 \leq K_c \leq 3 $~[9]. As we will see, our results for Boolean irreducibility corrections at $ p = 1/2 $ are well inside the uncertainty range. Furthermore we have corroborated our findings with new, and more precise simulations, due to the improvement of computational power since the 80's. Our results are in good agreement with the theoretical results.

The article is organized as follows: In Sec.~2 we write a combinatorial formulation for the dynamical equations (3), which allows us to work in a more suitable frame for calculations. In Sec.~3 the concept of Boolean irreducibility is introduced, and quantitative expressions for the number of functions with a fixed degree of irreducibility are obtained. In Sec.~4 a mean field approach, taking into account Boolean irreducibility is established, and the equation for the critical curve in phase space $ \left( p_c, K_c \right) $ is developed. In Sec.~5 the average of the degree of irreducibility times the probability of change of a Boolean function is calculated. This allows us, finally, to obtain the corrected-for-irreducibility critical curve $ \left( p_c, K_c \right) $, and compare it with Derrida's {\it et al.} result (6). In Sec. 6 we report our computer simulations results, which are in good agreement with our analytical results. In Sec.~7 we set up our conclusions. In the appendixes, some properties of the combinatorial coefficients which are used in the work are quoted.

\bigskip

\section{2. Combinatorial Notation for \textit{NK}-Automata}

Let us write the evolution equation (3) in a formal language elaborated in  Refs.~[3,4] that is more suitable for the understanding of the combinatorial structure of the \textit{NK}-Kauffman networks. Throughout the article: \break $ \forall \ S, S' \in \mathbb{Z}_2 $, $ S \oplus S' \in \mathbb{Z}_2 $ is intended to be addition {\it modulo $2$}. By $ \left[ N \right] = \left\{ 1, 2, \dots, N \right\} $ we denote the set of the first $ N $ natural numbers.

\begin{itemize}

\item[] {\bf Definition~1}:

Let
$$
\mathfrak{C}^N_K = \left\{ C_K^{(\alpha)} \right\}_{\alpha = 1, \dots, {N \choose K}}
$$
denote the collection of all the subsets of $ \left[ N \right] $ with cardinality $ K $ \break ($ 0 \leq K \leq N $), arranged in some unspecified order $ \alpha $.

\item[] {\bf Definition~2}:

Each element $ C_K^{(\alpha)} \in \mathfrak{C}^N_K $ is called a {\it $K$-connection set}, and is denoted by
$$
C_K^{(\alpha)} = \left\{ i_1, i_2, \dots, i_K \right\} \subseteq
\left[ N \right],
$$
with, $ i_1 < i_2 < \dots < i_K $; $ i_m \in \left[ N \right] $ ($ m = 1, \dots, K $).

\item[] {\bf Definition~3}:

To each $K$-connection set $ C_K^{(\alpha)} $ we associate a {\it $K$-connection function}
$$
C_K^{* (\alpha)}: \mathbb{Z}_2^N \longrightarrow \ \mathbb{Z}_2^K  \eqno(7)
$$
defined by
$$
C_K^{* (\alpha)} \left( {\bf S} \right) = C_K^{* (\alpha)} \left(
S_1, \dots, S_N \right) = \left( S_{i_1}, \dots, S_{i_K} \right) \ \ \forall \ {\bf S} \in \mathbb{Z}_2^N.
$$

\item[] {\bf Definition~4}:

A {\it $K$-Boolean function} is a map
$$
b_K: \mathbb{Z}_2^K \to \mathbb{Z}_2, \eqno(8)
$$
and its negation $ \neg b_K $ is given by $ \neg b_K = b_K \oplus 1 $.

\item[] {\bf Definition~5}:

A $K$-Boolean function (8) is completely determined by its {\it truth table} $ \mathfrak{B} \left( b_K \right) $, given by
$$
\mathfrak{B} \left( b_K \right) = \left[ \sigma_1, \sigma_2, \dots, \sigma_{2^K}
\right], \eqno(9)
$$
where, $ \sigma_s \in \mathbb{Z}_2 $, is the $s$-th image of (8) given by
$$
s = s \left( {\bf S} \right) = 1 + \sum_{i=1}^K \ S_i \ 2^{i-1} \hskip1.0cm 1 \leq s \leq 2^K , \eqno(10)
$$
which defines a {\bf total order} among the possible $ 2^K $ inputs of the argument $ {\bf S} \in \mathbb{Z}_2^K $ of the $K$-Boolean function (8).

There are $2^{2^K}$ $K$-truth tables $ \mathfrak{B} \left( b_K \right) $. Each $K$-Boolean function can be, uniquely classified according to Wolfram's notation by an integer number $ \mu = 1, \dots, 2^{2^K} $ given by~[4,16]
$$
\mu = 1 + \sum_{s=1}^{2^K} 2^{s-1} \sigma_s ;
$$
that also defines a {\bf total order} among the $K$-Boolean functions. So we add a superscript $ \mu $ to each of the $K$-Boolean functions (8) and make

\end{itemize}

\begin{itemize}

\item[] {\bf Definition~6}:

The set of {\it all $K$-Boolean functions} is given by
$$
\Xi_K = \left\{ b^{(\mu)}_K : \mathbb{Z}_2^K \longrightarrow \mathbb{Z}_2 \right\}_{\mu = 1}^{2^{2^K}}. \eqno(11)
$$

\end{itemize}


We clarify our abstract notation by the all important example of the truth table (9), for the $ K = 2 $ case; presented in Table~{\it 1}. The first line represents their Wolfram's number $ \mu $, indicated in boldface. At the bottom of the table: $ \mathfrak{F} $ stands for the logical meaning of each $2$-Boolean function, with $S_i$ ($i=1,2$) representing the identity $2$-Boolean function in the $i$-th argument of (10), while $ \neg S_i = S_i \oplus 1 $ its negation. The parameters $ \lambda $ and $ \omega $, to be defined below, represent: the degree of irreducibility, of the corresponding $2$-Boolean function, and its weight; respectively.


\footnotesize
\begin{center}
\begin{tabular}{|c|c|c|c|c|c|c|c|c|c|c|c|c|c|c|c|c|}
\hline $ \mathfrak{B} \left( b_2 \right) $ & {\bf 1} & {\bf 2}
& {\bf 3} & {\bf 4} & {\bf 5} & {\bf 6} & {\bf 7} & {\bf 8} & {\bf
9} & {\bf 10} & {\bf
11} & {\bf 12} & {\bf 13} & {\bf 14} & {\bf 15} & {\bf 16} \\
\hline $ \sigma_1  $ & 0 & 1 & 0 & 1 & 0 & 1 & 0 & 1 & 0 & 1 & 0 & 1 & 0 & 1 & 0 & 1 \\
\hline $ \sigma_2  $ & 0 & 0 & 1 & 1 & 0 & 0 & 1 & 1 & 0 & 0 & 1 & 1 & 0 & 0 & 1 & 1 \\
\hline $ \sigma_3  $ & 0 & 0 & 0 & 0 & 1 & 1 & 1 & 1 & 0 & 0 & 0 & 0 & 1 & 1 & 1 & 1 \\
\hline $ \sigma_4  $ & 0 & 0 & 0 & 0 & 0 & 0 & 0 & 0 & 1 & 1 & 1 & 1 & 1 & 1 & 1 & 1 \\ \hline $-$ & $-$ & $-$ & $-$ & $-$ & $-$ & $-$ & $-$ & $-$ & $-$ & $-$ & $-$ & $-$ & $-$ & $-$ & $-$ & $-$ \\
\hline $ {\mathfrak F} $ & $ \neg \tau $  & $\neg \vee$ & $ \nRightarrow $  & $ \neg S_2 $  & $ \nLeftarrow $ & $ \neg S_1 $ & $ \nLeftrightarrow $ & $ \neg \wedge $ & $ \wedge $  & $ \Leftrightarrow $ & $ S_1 $ & $ \Leftarrow $ & $ S_2 $ & $ \Rightarrow $ & $ \vee $ & $ \tau $  \\
\hline $ \lambda  $ & 0 & 2 & 2 & 1 & 2 & 1 & 2 & 2 & 2 & 2 & 1 & 2 & 1 & 2 & 2 & 0 \\ \hline $ \omega  $ & 0 & 1 & 1 & 2 & 1 & 2 & 2 & 3 & 1 & 2 & 2 & 3 & 2 & 3 & 3 & 4 \\
\hline

\end{tabular}
\end{center}
\normalsize \baselineskip = 12.4pt

\centerline{\textbf{Table~{\it 1}.} The $ \mathfrak{B} \left( b_2 \right) $ truth tables of the sixteen $2$-Boolean functions.}

For each Boolean variable $ S_i $, the maps (7) and (8) may be composed to represent map (3) in the following way
$$
\mathbb{Z}_2^N \ \stackrel{\ C_K^{*(\alpha_i)}}{\longrightarrow} \ \mathbb{Z}_2^K \
\stackrel{\ b_K^{(\mu_i)}}{\longrightarrow} \ \mathbb{Z}_2 \hskip1.0cm i = 1, \dots, N;
$$
with $ b_K^{(\mu_i)} $ and $ C_K^{* (\alpha_i)} $ extracted randomly according to the rules (1) and (2) respectively. Then, the evolution equation (3) may be rewritten as
$$
S_i \left( t + 1 \right) = b_K^{(\mu_i)} \circ C_K^{* (\alpha_i)}
\left( {\bf S} \left( t \right) \right), \ \ i = 1, \dots, N ,
$$
which defines an endomorphism $ \mathbb{Z}_2^N \to \mathbb{Z}_2^N $; but a particular one, due to the presence of the $K$-connection map $ C_K^{* (\alpha_i)} $. So, for $ K < N $; $ {\cal L}^N_K  \varsubsetneq {\cal B}_N \cong {\cal G}_{2^N} $, and only for the case $ K = N $; $ {\cal L}^N_N \equiv {\cal B}_N \cong {\cal G}_{2^N} $~[3,4].

The random construction of $ b_K^{(\mu_i)} $ and $ C_K^{*(\alpha_i)} $ is done in the following way:

\begin{itemize}

\item[{\it i})] According to (2): Extracting each function $ C_K^{*(\alpha_i)} $, with equiprobability and without repetition, among the possible $ {N \choose K} $ $K$-connection sets.

\item[{\it ii})] According to (1): Extracting each $ b_K^{(\mu_i)} $ from the probability distribution
$$
\Pi_p \left( b_K \right) = p^\omega \left( 1 - p \right)^{2^K - \,
\omega} \ , \eqno(12)
$$
 such that in the truth table (9) $ \sigma_s = 1 $, ($ s = 1, \dots, 2^K $) with probability $ p $, and $ \sigma_s = 0 $ with probability $ 1 - p $; and where $ \omega = 0, 1, \dots, 2^K $ denotes the value of the {\it weight function} $ \omega \left( b_K \right) $ of $ b_K $, defined by
$$
\omega \left( b_K \right) = \sum_{s=1}^{2^K} \sigma_s. \eqno(13)
$$
The following decomposition of $ \Xi_K $ is going to be important for the calculations of the next section:
$$
\Xi_K = \bigsqcup_{\omega = 0}^{2^K} \  \mathfrak{P}_K \left( \omega \right) , \eqno(14.a)
$$
where
$$
\mathfrak{P}_K \left( \omega \right) = \left\{ b_K \in \Xi_K | \, \omega \left( b_K \right) = \omega \right\} \eqno(14.b)
$$
with cardinality
$$
\# \mathfrak{P}_K \left( \omega \right) = {2^K \choose \omega}. \eqno(14.c)
$$

\end{itemize}

\section{3. Irreducible Boolean Functions}

In fact, not all of the $K$-Boolean functions depend strictly on their $K$ arguments. For example, for $ K = 2 $, in Table~{\it 1}: Rules {\bf 1} and {\bf 16} ({\it contradiction} and {\it tautology}, respectively) do not depend on either $ S_1 $ or $ S_2 $;  rules {\bf 4} {\bf 6}, {\bf 11}, and {\bf 13} depend only on one of their arguments; while the remaining $10$ depend on both $ S_1, S_2 $. Due to these facts, we do the following


\

\leftline{{\bf Definitions 7}}

\begin{itemize}

\item[{\it i})] A $K$-Boolean function $ b_K $ is {\it irreducible} on its $m$-th argument $ S_m $ ($ m = 1, \dots, K $), iff there exists an $ {\bf S} \in \mathbb{Z}_2^K $ for which
$$
b_K \left( S_1, \dots, S_m, \dots, S_K \right) = 1 \oplus b_K \left( S_1,
\dots, S_m \oplus 1, \dots, S_K \right),
$$
while, if this does not happen, the $K$-Boolean function $ b_K $ is {\it reducible} on the $m$-th argument $ S_m $.

\item[{\it ii})] A $K$-Boolean function $ b_K $ is said to have a {\it degree of irreducibility} $ \lambda $ ($ \lambda = 0, 1, \dots, K $); if it is irreducible on $ \lambda $ of their arguments and reducible on the remaining $ K - \lambda $.

\item[{\it iii})] If $ \lambda = K $, the $K$-Boolean function is called {\it totally irreducible}.

\end{itemize}

Let us denote by $ \lambda \left( b_K \right) $  the function that gives the {\it degree of irreducibility} of $ b_K $, and $ \lambda $ ($ 0 \leq \lambda \leq K $) their possible values. Then $ \Xi_K $ may also be decomposed like
$$
\Xi_K = \bigsqcup_{\lambda = 0}^K \ {\mathfrak T}_K \left( \lambda
\right), \eqno(15)
$$
where
$$
{\mathfrak T}_K \left( \lambda \right) = \left\{ b_K \in \Xi_K | \, \lambda\left( b_K \right) = \lambda \right\} .
$$
The cardinal coefficients $ \beta_K \left( \lambda \right) \equiv \# {\mathfrak T}_K \left( \lambda \right) $ were calculated recursively, in Ref.~[4] obtaining the formula
$$
\beta_K \left( \lambda \right) = {K \choose \lambda} \
\mathfrak{G}_\lambda , \eqno(16)
$$
where $ \mathfrak{G}_\lambda \equiv \beta_\lambda \left( \lambda \right) $ is the number of totally irreducible $\lambda$-Boolean functions. From (11), and taking cardinalities in (15); it happens that (16) obeys the formula
$$
2^{2^\lambda} = \sum_{\lambda = 0}^K {K \choose \lambda} \
\mathfrak{G}_\lambda ;
$$
which may be inverted using Comtet's formulas~[17] for the combinatorial coefficients (see Appendix~{\bf A}) obtaining
$$
\mathfrak{G}_\lambda = \sum_{m = 0}^\lambda \left( -1 \right)^{\lambda - m} \, {\lambda \choose m} \ 2^{2^m} . \eqno(17)
$$
All coefficients $ \beta_K \left( \lambda \right) $, but $ \beta_K \left( 0 \right) = 2 $, grow with $K$. $ {\mathfrak T}_K \left( 0 \right) $ consists of the $K$-{\it contradiction} $ \neg \tau \equiv b_K^{(1)} $ and $K$-{\it tautology} $ \tau \equiv b_K^{(2^{2^K})} $ functions, with truth tables (9) given by
$$
\mathfrak{B} \left( \neg \tau \right) = [ \underbrace{0, 0, \dots, 0}_{2^K} ],  \eqno(18)
$$
and
$$
\mathfrak{B} \left( \tau \right) = [ \underbrace{1, 1, \dots, 1}_{2^K} ]  \eqno(19)
$$
respectively.

On the other extreme, from (16) and (17) we obtain for the number of totally irreducible functions $ \beta_K \left( K \right) $ the asymptotic expression,
$$
{\beta_K \left( K \right) \over 2^{2^K}} \approx 1 - {\cal O} \left( {K \over 2^{2^{K-1}}}\right) , \eqno(20)
$$
for $ K \gg 1 $. This shows that, with respect to the normalized counting measure, almost any $K$-Boolean function is totally irreducible.

These facts show us that the ``real connectivity" of a $K$-Boolean function $ b_K $ is not $ K $, but $ \lambda \left( b_K \right) $. However, for big values of $ K $ the ``real connectivity" becomes nearly $ K $. Note that the curve for the phase transition given by (6), in the region $ p \sim 1 / 2 $ predicts values of the order $ K \sim 2 $. So, for this region the effect of irreducibility should be appreciable in a mean field treatment that takes into account the degree of irreducibility due to the small values of $ K $ there. Let us calculate the average value $ \left< \lambda \right> $ as a function of  $ p $ and $ K $. This represents the average connectivity of $ b_K $ with respect to the extraction probability (12). So
\begin{eqnarray*}
\left< \lambda \right> &=& \sum_{b_K \in \Xi_K} \ \lambda \left( b_K \right) \, \Pi_p \circ \omega \left( b_K \right) = \sum_{\lambda = 0}^K \ \lambda \sum_{b_K \in {\mathfrak T}_K \left( \lambda \right)} \ \Pi_p \circ \omega \left( b_K \right) \\ &=& \sum_{\lambda = 0}^K \ \lambda \ \sum_{\omega = 0}^{2^K} \Pi_p \left(\omega \right) \sum_{b_K \in \left[{\mathfrak T}_K \left( \lambda \right)\cap \mathfrak{P}_K \left( \omega \right) \right]} 1 \\ &=& \sum_{\lambda = 0}^K \ \lambda \ \sum_{\omega = 0}^{2^K} \Pi_p \left(\omega \right) \, \varrho_K \left( \lambda, \omega \right) , \hskip6.1cm (21)
\end{eqnarray*}
where
$$
\varrho_K \left( \lambda, \omega \right) = \# \left[{\mathfrak T}_K \left( \lambda \right)\cap \mathfrak{P}_K \left( \omega \right) \right] .
$$
The calculation of the cardinality $ \varrho_K \left( \lambda, \omega \right) $ is a difficult task that has been done in Ref.~[14] using algebraic theoretical tools to do the combinatorial counting. Here we quote the result and refer the interested reader to the bibliography~[14]:
\begin{eqnarray*}
\varrho_K \left( \lambda, \omega \right) &=& {K \choose \lambda} \ \sum_{m=0}^\lambda \left( -1 \right)^{\lambda - m} {\lambda \choose m} \\ \, & \times & \delta \left( \left\lfloor \omega \, 2^{m-K} \right\rfloor - \omega \, 2^{m-K} \right) \ \ {2^m \choose \left\lfloor \omega \, 2^{m-K} \right\rfloor} ,
\end{eqnarray*}
where for all $ a \in \mathbb{R} $,

$$
\delta \left( a \right) = \left\{ \begin{array}{ll}
1 & \mbox{if $ a = 0 $}
\\ {} & {} \\ 0 & \mbox{if $ a \neq 0 $} \end{array} \right.
$$
is Kronecker's delta, $ \lfloor a \rfloor \in \mathbb{Z} $ the {\it floor function}, which denotes the greatest integer $ \lfloor a \rfloor $ such that $ \lfloor a \rfloor \leq a $, $ {\lambda \choose m} = 0 $, for $ m > \lambda $, and $ 0^0 \equiv 1 $. With this aid, (21) may be calculated, obtaining
$$
\left< \lambda \right> = K \, \left( 1 - \left[ 1 - 2 p \left( 1 - p \right) \right]^{2^{K-1}} \right) < K . \eqno(22)
$$
See Appendix~{\bf B} for manipulation of the combinatorial coefficients, and representations of Kronecker's delta in terms of them.

For $ p \sim 1/2 $, the effect of (20) tends to dominate for $ K \gg 1 $, making $ \left< \lambda \right> \sim K $. Instead, for $ p $ near to $0$ or $1$, the $K$-Boolean function (18) or (19), respectively dominates (since contradiction and tautology are, for each case, the only functions to have a significant probability to be extracted); thus making $ \left< \lambda \right> \sim 0 $. In Fig.~1, a graph of $ \left< \lambda \right> / K $ {\it vs.} $ p $ shows this behavior for different constant values of $ K $.

\section{4. Mean Field Theory for \textit{NK}-Automata}

Now a mean field approach ($ N \gg 1 $) is developed  to study the behavior of the Hamming distance of two initially nearby states $ {\bf S} $ and $ {\bf S'} $, with respect to the parameters $ K $, and $ p $ of the \textit{NK}-Automata.

\begin{itemize}


\item[] {\bf Definition~8}:

The {\it Hamming distance} $ d_H $ between two states $ {\bf S}, {\bf S'} \in \mathbb{Z}_2^N $ is given by
$$
d_H \left( {\bf S}, {\bf S'} \right) = \sum_{i=1}^N \left( S_i \oplus S'_i \right) . \eqno(23)
$$

\end{itemize}

We want to see the behavior of (23) as the system evolves in time according to (3) starting at $ t = 0 $ with two arbitrary states $ {\bf S} (0) $, $ {\bf S'} (0) $, which are nearby (in relation to $ N $), that is $ d_H \left( {\bf S}(0), {\bf S'}(0) \right) \ll N $. Let us use the shorthand notation
$$
d_H(t) \equiv \sum_{i=1}^N \left( S_i (t) \oplus S'_i (t) \right)
$$
for the Hamming distance of the evolving states $ {\bf S}(0) $ and $ {\bf S}'(0) $ at time $t$. Since $ 0 \leq d_H(t) \leq N $: without loss of generality we may write
$$
d_H(0) \equiv \varepsilon N \ , \hskip0.5cm 0 < \varepsilon \ll 1 \eqno(24)
$$
for the initial Hamming distance, where $ \varepsilon $ is a fixed value not depending on $ N $. Then $ 1 \ll d_H(0) \ll N $ and so, for $ N \gg 1 $, statistics may be done. Since the Boolean functions and their connections are randomly chosen from (1), and (2), respectively; we have that:

Each affected site $j$, such that $ S_j(0) \neq S'_j(0) $; will affect, on average, $K$ sites $ C_K \equiv \left\{ j_1, \dots, j_K \right\} \subseteq \left[ N \right] $. The $j_l$-th affected site ($l = 1, \dots, K $) is the argument of a $K$-Boolean function $ b_K^{(i)} $ which is a stochastic variable obtained from the probability distribution (12). So, $ b_K^{(i)} $ is going to have a degree of irreducibility $ \lambda \left( b_K^{(i)} \right) $, and a {\it probability of change} $ P_\chi \left( b_K^{(i)} \right) $ owing that one of their arguments has changed (to be calculated in the next section). We take as a mean field approximation, that in average the Hamming distance will increase or decrease, for each site $ i $ by a factor $ 0 \leq \Delta \left( K, p \right) \leq K $ at each time step. Then
$$
d_H \left( 1 \right) = d_H \left( 0 \right) \, \Delta \left( K, p \right) =  \varepsilon N \ \Delta \left( K, p \right) \ ,
$$
where
$$
\Delta \left( K, p \right) = {1 \over \varepsilon N} \sum_{i=1}^{\varepsilon N} \, P_\chi \left( b_K^{(i)} \right) \, \lambda \left( b_K^{(i)} \right). \eqno(25)
$$
Since the stochastic terms in the sum are statistically independent we may apply the central limit theorem for $ N \gg 1 $ to obtain
$$
\Delta \left( K, p \right) \approx \left< P_\chi \left( b_K \right) \, \lambda \left( b_K \right) \right> = \sum_{b_K \in \Xi_K} \Pi_p \left( b_K \right) \, P_\chi \left( b_K \right) \, \lambda \left( b_K \right).
$$

The same arguments are valid for any $t$, as long as $ 1 \ll d_H(t) \ll N $, continues to be true, so we obtain
$$
d_H \left( t + 1 \right) = d_H \left( t \right) \, \left< \lambda \left( b_K \right) \, P_\chi \left( b_K \right) \right> . \eqno(26)
$$
The {\it relative error} $ \mathcal{E}_r (t) $ in the calculation of (26) can by estimated in terms of the variance $ \Sigma^2 (t) $ of $ d_H \left( t \right) $. Using the central limit theorem once again we have $ \Sigma^2 \approx d_H \left( t \right) \, \varsigma^2 $, where $ 0 \leq \varsigma^2 \leq K^2 $. So we obtain
$$
\mathcal{E}_r (t) \equiv {\Sigma (t) \over d_H \left( t \right)} \approx {\varsigma \over \left< \lambda \left( b_K \right) \, P_\chi \left( b_K \right) \right> \, \sqrt{d_H \left( t \right)}} \sim {\cal O} \left( {1 \over \sqrt {N}} \right) \ . \eqno(27)
$$
Since $ \mathcal{E}_r (t) $ vanishes for $ N \to \infty $, the mean field approximation is exact in the thermodynamic limit.

Solving for the initial condition $ d_H(0) = \varepsilon \, N $, we have for the evolution of Hamming distance (26) the mean field equation
$$
d_H (t) = \varepsilon \, N \ \exp \left\{ t \ \ln \left[ \Delta \left( K, p \right) \right] \right\}.
$$
Now we see that there is an exponential grow (or decay) in $ d_H (t) $ depending on whether $ \Delta \left( K, p \right) $ is bigger (or smaller) than one: this divides the phase space of the parameters $ p $ and $ K $ in the regions

$$
\Delta \left( K, p \right) = \left\{ \begin{array}{ll}
>  1 & \mbox{Standing for a disordered phase, called {\it chaotic},}
\\ {} & {} \\ < 1 & \mbox{Representing an {\it ordered}, or {\it frozen phase},} \end{array} \right.
$$
while
$$
\Delta \left( K_c, p_c \right) = 1  \eqno(28)
$$
represents the equation for the critical transition curve.


\section{5. Phase Space Diagram corrected for Boolean Decomposition}

We now study for which values of the parameters $ K $, and $ p $ (28) holds. This is done calculating the average $ \left< \lambda \left( b_K \right) \, P_\chi \left( b_K \right) \right> $. Note aboard that, due to the fact that $ 0 \leq \left< \lambda \left( b_K \right) \right> < K $, and $ 0 \leq \left< P_\chi \left( b_K \right) \right> \leq 1 $: $ \Delta \left( K, p \right) < 1 $ for $ K \leq 1 $. So $ K > 1 $ is a necessary, but not sufficient, condition for chaotic behavior to be exhibited in \textit{NK}-Kauffman networks.

The probability that a $K$-Boolean function $ b_K $ changes; due that one of its arguments has changed $ P_\chi \left( b_K \right) $ is given by definition as,
$$
P_\chi \left( b_K \right) = \sum_{\sigma \in \mathbb{Z}_2} \pi \left( b_K: \sigma \right) \pi \left( b_K: \sigma \oplus 1 \ | \ \sigma \right). \eqno(29)
$$
Where $ \pi \left( b_K: \sigma \right) $ is the probability to extract at random the value $ \sigma \in \mathbb{Z}_2 $ from the truth table (9) $ \mathfrak{B} \left( b_K \right) $, and $ \pi \left( b_K: \sigma \oplus 1 \ | \ \sigma \right) $ is the probability to extract at random the value $ \neg \sigma \equiv \sigma \oplus 1 $, from $ \mathfrak{B} \left( b_K \right) $; given that $ \sigma $ has been previously extracted. Then from (13)
$$
\pi \left( b_K: \sigma \right) = {\omega \left( b_K \right) \over 2^K} \ \delta \left( \sigma \oplus1 \right) + {2^K - \omega \left( b_K \right) \over 2^K} \ \delta \left( \sigma \right) .
$$
While
$$
\pi \left( b_K: \sigma \oplus 1 \ | \ \sigma \right) = \Sigma_K \ \pi \left( b_K: \sigma \oplus 1 \right) ,
$$
where
$$
\Sigma_K \equiv {2^K \over 2^K - 1}
$$
is a {\it second extraction factor} which appears since now there remain in the pool $ 2^K -1 $ states $ {\bf S} \in \mathbb{Z}_2^N $ to choose. Substituting in (29) we obtain
$$
P_\chi \left( b_K \right) \equiv P_\chi \circ \omega \left( b_K \right)= 2 \ \Sigma_K  \ {\omega \left( b_K \right) \left( 2^K - \omega \left( b_K \right) \right) \over 2^{2K}} .
$$
Now we may calculate $ \Delta \left( K, p \right) = \left< \lambda \left( b_K \right) \ P_\chi \left( b_K \right) \right> $ in the same way as (22) through (21):
\begin{eqnarray*}
\Delta \left( K, p \right) &=& \sum_{b_K \in \Xi_K} \, \Pi_p \circ \omega \left( b_K \right) \, \lambda \left( b_K \right) \, P_\chi \circ \omega \left( b_K \right) \\ &=& \sum_{ \omega = 0}^{2^K} \, \Pi_p \left( \omega \right) \, P_\chi \left( \omega \right) \, \sum_{\lambda = 0}^K \ \lambda \ \varrho_K \left( \lambda, \omega \right)  .
\end{eqnarray*}
Which with the aid of (B3) and (B5) of Appendix~{\bf B}
$$
\Delta \left( K, p \right) =  K \, 2 \, p \left( 1 - p \right) \left\{ 1 - 2 \, p \, \left( 1 - p \right) \left[ 1 - 2 \, p \, \left( 1 - p \right) \right]^{2^{K-1}-2} \right\}
$$
is obtained.

So the critical transition curve (28) is given by
$$
K_c \, 2 \, p_c \left( 1 - p_c \right) \left\{ 1 - 2 \, p_c \, \left( 1 - p_c \right) \left[ 1 - 2 \, p_c \, \left( 1 - p_c \right) \right]^{2^{K_c - 1}-2} \right\} = 1 . \eqno(30)
$$
Comparison with result (6) shows the appearance of a new factor $ \left\{ \cdots \right\} $ of order $ \mathcal{O} \left[ 1 -  2 \, p_c \left( 1 - p_c \right) \right] $ which accounts for the existence of Boolean irreducibility in the functions. Note that $ \left\{ \cdots \right\} $ appears since Derrida {\it et al.}, not taking account for Boolean irreducibility implicitly calculated~[9,11,12]:
$$
K_c \ \left<  P_\chi \left( b_K \right) \right> = \sum_{b_K \in \Xi_K} \, \Pi_p \circ \omega \left( b_K \right) \, P_\chi \circ \omega \left( b_K \right) = K_c \ 2 \, p_c \left( 1 - p_c \right) = 1 \ .
$$

Fig.~{\it 2} compares the graphs of (6) and (30). Now, due to the effect of irreducibility the transition occurs for each $ p $ at greater values of $ K $. In particular, for the all important case $ p = 1 / 2 $;
$$
K_c = 2.62140224613 \dots . \eqno(31)
$$

\section{6. Simulations}

We have done numerical simulations for $ N = 10^6 $ automata. Since a mean field approximation implies $ N \to \infty $, we should expect the appearance of non ideal results, which come from the fact that the relative error (27) is not zero.

First of all, since we use extensively random numbers, we use a pseudo-random number generator routine with a long period ($ > 2 \times 10^{18} $) which generates numbers from a uniform distribution on the open interval $ \left( 0, 1 \right) $~[18]. The study is done fixing  $ K = 3 $, and varying $ p $ from the predicted value by Derrida {\it et al.} for the phase transition (6) $ p = 0.2113\ldots $. Then we start changing values from $ p = 0.22, 0.23 $ through the predicted value (30) by taking into account Boolean irreducibility $ p = 0.2654\ldots $ Finally, we go to the chaotic phase values $ p = 0.3 $ and $ p = 0.5 $.

As a double check, we also consider the special case $ p = 0.5 $ for the controversial connectivity value $ K = 2 $.

The flow of the program is the following:

\begin{itemize}

\item[${[A]}$] The routine generates an automata using the random rules (1) and (2), then two random initial states are generated with the restriction that their Hamming distance be $ d_H (0) = 10^4 $. The dynamical system is then ran $ 29 $ {\it time} steps. We repeat this process starting with different random initial states and ran the process $ 10 $ times, then the average is taken. This step is done in order to avoid be fooled in particular initial states.

\item[${[B]}$] A new Kauffman's automata is constructed (same $ K = 3 $, and $ p $) and step [\textsl{A}] is repeated.

\item[${[C]}$] Steps [\textsl{A}] and [\textsl{B}] are leave running $100$ times. Once ended, the average and the standard deviation are calculated.

\item[${[D]}$] The results are stored in files where graphs for $ d_H (t) $ {\it vs.} $ t $ and the error bars for $ d_H (t) $, obtained from the standard deviation, are depicted.

\end{itemize}

Before we proceed to interpret the resulting graphs, some explanations must be considered: $ d_H (0) = 10^4 $ is taken big enough to make the relative error $ {\cal E}_r (0) $ (27) as small as possible, while avoiding saturation effects due to the finite value of $ N = 10^6 $. The dynamical system is then leave running $ 29 $ {\it time} steps, which is more than enough, in order that Hamming distance does not become too saturated. This effect happens when Hamming distance stops growing, or decreasing exponentially having an inflexion toward a constant value. Note that $ \varepsilon $ in (24) is fixed, so $ d_H (0) = \varepsilon \, N $ scales like $ N $.

In Fig.~3 we see the graph of $ d_H (t) $ {\it vs.} $ t $ for the case $ p = 0.2113\ldots $ ($ K = 3 $). We may observe a clear decay of the graph showing that the phase transition does not occur there as predicted by (6) in which irreducibility was not taken into account. In Fig.~4 we observe the graphs for the cases $ p = 0.22 $, $ p = 0.23 $, and $ p = 0.2654\ldots $ which are in the critical zone. Apparently the phase transition does not occur at the predicted value (corrected for irreducibility) $ p = 0.2654\ldots $, which is growing instead of being neither growing, nor decaying. Furthermore the value $ p = 0.22 $ has a lower slope, but is still growing. This is an artefact due to the finite character of $ N $. From (25) applying the central limit theorem, we obtain
$$
\Delta_{t} \left( K, p \right) \simeq \Delta \left( K, p \right) \pm {\theta \over \sqrt{d_H (t)}} \ , \eqno(32)
$$
where $ \theta^2 $ is the variance of each term in the summation (25), and we have set $ d_H(t) $ instead of $ \varepsilon \, N $ in (32). So for $ N $ finite there is not a phase transition curve (a fact well known in statistical mechanics) but a region whose thick goes to zero like $ 1 / \sqrt{N} $. In Fig.~5 we see the graphs for $ p = 0.3 $ and $ p = 0.5 $ which are clear in the chaotic region. Their initial exponential increment is evident, and their saturation starts at about $ t \sim 10 $.

We ran also  a simulation for the special case $ K = 2 $ at $ p = 1 / 2 $ the graph is shown in Fig.~6 which clearly shows that the automata are in the ordered phase zone. The increase of the error bars is due to the decrease of $ d_H(t) $ which increases the value of the relative error (27). So, since we have shown numerically that for $ p = 0.5 $, $ K = 3 $ the automata are in a disordered phase; the transition must occur well inside $ 2 < K < 3 $, with a theoretical predicted value given by (31), when irreducibility is taken into account.

We have run simulations for $ d_H (0) = 10^5 $. For $ N = 10^6 $ this value of $ d_H (0) $ is very near the maximum possible value of Hamming distance; {\it i.e.} $ N = 10^6 $. However it still gives information. Since $ d_H (0) $ is bigger, errors (27) and (32) are smaller; and we could expect a better approximation to the transition value. We leaved ran the dynamical system only $ 9 $ {\it time} steps since Hamming distance saturates very fast. The results are reported in Fig.~7 for the near to the critical phase transition values $ p = 0.2113\ldots, \, 0.22, \, 0.23, \, 0.24 $ and $ 0.2654 \ldots $ One observes that the slopes of all the curves have decreased with only the curve for $ p = 0.2654 \ldots $ having a positive slope. Note also that the error bars have decreased. Once again this is due to a decrease of errors (27) and (32).

We may conclude with certainty that numerical simulations with $ d_H (0) = \varepsilon \, N $ and $ N > 10^6 $ will approach further to the theoretical predicted value for the phase transition $ p = 0.2654 \ldots $ for $ K = 3 $ accordingly to (30).

\newpage

\section{7. Conclusions}

We have re-calculated the phase transition curve (6) for \textit{NK}-Kauffman networks~[11,12] correcting calculations for the effect of Boolean irreducibility of $K$-Boolean functions~[4]. While it turns out not to be a big correction for this case, it has an important effect in many other aspects of the behavior of \textit{NK}-Kauffman networks, such as the injective properties of the function $ \Psi $ Eq.~(4), which maps the \textit{NK}-Kauffman networks set $ \mathcal{L}_K^N $ into the $ 2^N $ functional graphs set $ \mathcal{G}_{2^N} $~[3,4]. Important to be noted is that result (31), for $ p = 1 / 2 $ and $ K = 2 $, is well inside the uncertainty $ 2 \leq K_c \leq 3 $ of past numerical simulations~[1,9,15]. We also ran simulations for $ K = 3 $ in the region near $ p = 0.2654\ldots $ [the value that (30) predicts for $ p_c $] obtaining good agreement with the mean field treatment which takes into account Boolean irreducibility. As a double check we ran simulations for $ p = 1 /2 $, $ K = 2 $. Our results clearly indicate that it corresponds to an ordered phase in accordance with the theoretical results of this work.

Without doubt the degree $ \lambda $ of irreducibility, {\it Definition 7 (ii)} should play an important role in the characterization of \textit{NK}-Kauffman networks dynamics as a function of parameters $ N $, $ K $, and $ p $. A possible line of research may be to try to correct for Boolean irreducibility the study made by Derrida \& Stauffer for Kauffman cellular automata in a two dimensional lattice~[11].

\section{Acknowledgments}

This work is supported in part by {\bf PAPIIT} projects Nos.~{\bf IN101309-3} and {\bf IN102712-3}. The author wish to thank: Alberto Verjovsky and Fabio Benatti for fruitful mathematical discussions. Thal\'\i a Figueras for careful reading of the manuscript, Pilar L\'opez Rico for accurate services on informatics, Mariana Zertuche for data reduction, V\'\i ctor Dom\'\i ngez and Fernando Gonz\'alez for computer's advise. Last, but not least, the author is indebted with second reviewer for his/her suggestions which contributed to improve article's quality.


\newpage

\section{Appendix A: Inversion Formula for Binomial Coefficients}

In Comtet's work, the following inversion formula is proved~[17]:

For any two sequences of real numbers
$$
\left\{ f_r \right\}_{r=0}^n, \ {\rm and} \ \left\{ g_r \right\}_{r=0}^n,  \ n \geq 0
$$
such that
$$
f_n = \sum_{r=0}^n {n \choose r} \, g_r .
$$
Then, it follows that the $ g_r $ are given in terms of the $ f_r $ through
$$
g_n = \sum_{r=0}^n \left( -1 \right)^{n-r} \, {n \choose r} \, f_r .
$$

\section{Appendix B: Identities and checks for manipulating $ \varrho_K \left( \lambda, \omega \right) $}

It is useful for the calculations involving $ \varrho_K \left( \lambda, \omega \right) $ to extend the definition of the combinatorial coefficients when the upper index $ a \in \mathbb{R} $, and the lower index $ n \in \mathbb{Z}$; by writing~[19]:
$$
{a \choose n} = \left\{ \begin{array}{ll}
{a \left( a - 1 \right) \dots \left( a - n + 1 \right) \over n!} & \hskip0.5cm \mbox{for $ n \geq 0 $}
\\ {} & {} \\ 0 & \hskip0.5cm \mbox{for $ n < 0 $} \end{array} \right. . \eqno(B1)
$$
Which for the case $ a \in \mathbb{Z} $ gives $ {a \choose n} = 0 $ if $ a < n $. From (B1) the following identity can be proved to hold for any $ a \in \mathbb{R} $, $ m, n \in \mathbb{Z} $~[19]
$$
{a \choose m} {m \choose n} = {a \choose n} {a - n \choose m - n} . \eqno(B2)
$$
The {\it Binomial Theorem} for $ z \in \mathbb{C} $ comes to be
$$
(1 + z)^a = \sum_{m \geq 0} \, {a \choose m} \ z^m \hskip.5cm {\rm for} \hskip.5cm \left\{ \begin{array}{ll}
\left| z \right| < 1 , & \hskip0.5cm \mbox{if $ a \in \mathbb{R} $}
\\ {} & {} \\ \forall z \in \mathbb{C} & \hskip0.5cm \mbox{if $a \in \mathbb{N} \cup \left\{ 0 \right\} $} \end{array} \right. \eqno(B3)
$$
where, $ 0^0 \equiv 1 $. From (B3) the following useful Kronecker's delta representations, for $ a \in \mathbb{R}^+ \cup \{ 0 \} $ may be obtained by deriving with respect to $ z $ and taking $ z = -1 $ [where it is to be noted that the series (B3) still converges for $ z = - 1 $ due to the alternating sign, and that $ {a \choose m} \sim {\cal O} \left( m ^{-1 -a} \right) $, for $ m \gg 1 $]~[19];
$$
\delta \left( a \right) = \sum_{m \geq 0} \, \left( -1 \right)^m \, {a \choose m} , \hskip0.4cm \delta \left( a - 1 \right) = \sum_{m \geq 0} \, \left( -1 \right)^{m+1} \, m \,  {a \choose m} \ \ a \geq 0 . \eqno(B4)
$$

Using (B2) and (B4) one easily obtains the following check identities for $ \varrho_K \left( \lambda, \omega \right) $, which are consequences of (14), (16) and (17):
$$
\sum_{\lambda=0}^K \ \varrho_K \left( \lambda, \omega \right) = {2^K \choose \omega} ,
$$
and
$$
\sum_{\omega=0}^{2^K} \ \varrho_K \left( \lambda, \omega \right) \ = \ {K \choose \lambda} \sum_{m = 0}^\lambda \left( -1 \right)^{m - \lambda} \, {\lambda \choose m} \ 2^{2^m} \ \equiv \ \beta_K \left( \lambda \right) .
$$
From (B3) and (B4) also follows the useful identity
$$
\sum_{\lambda=0}^K \, \lambda \, \varrho_K \left( \lambda, \omega \right) = K \ \left[ {2^K \choose \omega} - {2^{K-1} \choose \lfloor {\omega \over 2} \rfloor } \ \delta \left( \lfloor {\omega \over 2} \rfloor - {\omega \over 2} \right) \right] \ . \eqno(B5)
$$

\

\newpage

{\bf References}

\begin{itemize}

\item[${[1]}$] S.A. Kauffman, {\it Metabolic Stability and
Epigenesis in Randomly Connected Nets}. J.~Theoret.~Biol. {\bf 22}
(1969) 437.

\item[${[2]}$] S.A. Kauffman, {\it The Origins of Order:
Self-Organization and Selection in Evolution}. Oxford University
Press (1993).

\item[${[3]}$] D. Romero, and F. Zertuche, {\it Number of
Different Binary Functions Generated by \textit{NK}-Kauffman
Networks and the Emergence of Genetic Robustness}. J.~Math.~Phys.
{\bf 48} (2007) 083506.

\item[${[4]}$] F. Zertuche, {\it On the robustness of NK-Kauffman networks against changes in their connections and Boolean functions}. J.~Math.~Phys.
{\bf 50} (2009) 043513.

\item[${[5]}$] D. Romero, and F. Zertuche, {\it The Asymptotic
Number of Attractors in the Random Map Model}.
J.~Phys.~A:~Math.~Gen. {\bf 36} (2003) 3691; {\it Grasping the
Connectivity of Random Functional Graphs}.
Stud. Sci. Math. Hung. {\bf 42} (2005) 1.

\item[${[6]}$] B. Derrida, and H. Flyvbjerg, {\it The Random
Map Model: a Disordered Model with Deterministic Dynamics}.
J.~Physique {\bf 48} (1987) 971.

\item[${[7]}$] H. Flyvbjerg, and N.J. Kjaer, {\it Exact
Solution of Kauffman's Model with Connectivity One}.
J.~Phys.~A:~Math.~Gen. {\bf 21} (1988) 1695.

\item[${[8]}$] J. Lynch, {\it On the threshold of chaos in random Boolean cellular automata}. Random Struct. Algorithms {\bf 6} (1995) 239-260; {\it Critical Points for Random Boolean Networks}. Physica~D {\bf 172} (2002) 49-64; {\it Dynamics of Random Boolean Networks}. In: Current Developments in Mathematical Biology, Proceedings of the Conference on Mathematical Biology and Dynamical Systems, 15--38. Eds.  K. Mahdavi, R. Culshaw, and J. Boucher. World Scientific Publishing Co. Publisher (2007).

\item[${[9]}$] B. Derrida, and  Y. Pomeau, {\it Random Networks of Automata: A Simple Annealed Approximation}. Europhys.~Lett. {\bf 1} (1986) 45.

\item[${[10]}$] B. Derrida, G. Weisbuch, {\it Evolution of overlaps between configurations in random Boolean networks}. J.~Physique {\bf 47} (1986) 1297.

\item[${[11]}$] B. Derrida, and D. Stauffer, {\it Phase Transitions in Two-Dimensional Cellular Automata.} Europhys.~Lett. {\bf 2} (1986) 739.

\item[${[12]}$] B. Derrida, {\it Dynamical Phase Transitions in Spin Models and Automata}. In: Fundamental Problems in Statistical Mechanics VII, 273--308. Ed. H. van Beijeren. Elsevier Science Publishers B.V., (1990).

\item[${[13]}$] M. Aldana, S. Coppersmith, and L. Kadanoff, {\it
Boolean Dynamics with Random Couplings}. In: Perspectives and Problems
in Nonlinear Science, 23--89. Springer Verlag, New York (2003).

\item[${[14]}$] M. Takane, and F. Zertuche, {\it $ \mathbb{Z}_2 $-Algebras in the Boolean Function Irreducible Decomposition}. J.~Math.~Phys. {\bf 53} (2012) 023516.

\item[${[15]}$] A.E. Gelfand, and C.C. Walker {\it Ensemble Modeling}. Marcel Dekker, Inc. New York and Basel (1984).

\item[${[16]}$] G. Weisbuch, {\it Complex Systems Dynamics}.
Addison Wesley, Redwood City, CA (1991); Wolfram, S., {\it
Universality and Complexity in Cellular Automata}. Physica~D {\bf
10} (1984) 1.

\item[${[17]}$] L. Comtet, {\it Advanced Combinatorics}. Reidel, (1974), p. 165.

\item[${[18]}$] W.H.Press, S.A. Teukolsky, W.T. Vetterling, and B.P. Flannery, {\it Numerical Recipes in C}. Second Ed. Cambridge, (1997).

\item[${[19]}$] R.L. Graham, D.E. Knuth, and O. Patashnik, {\it Concrete Mathematics}, New York: Addison-Wesley (1994).

\end{itemize}

\newpage

\centerline{\bf Figure caption}

\

\noindent
Figure~1. (Color online), Graph of $ \left< \lambda \right> / K $ {\it vs.} $ p $, for constant values of $ K $. It shows the behavior of the {\it average degree of irreducibility} of the $K$-Boolean functions $ b_K $ {\it vs.} the bias $p$.

\

\noindent
Figure~2. (Color online) Compares the graphs of phase transition curve (6), calculated by Derrida, {\it et al.}, which does not take into account the irreducible degree of $K$-Boolean functions, with the graph of (30) which takes into account this effect.

\

\noindent
Figure~3. (Color online)  The graph of $ d_H(t) $ {\it vs.} $ t $ for the parameters values $ K = 3 $ and $ p = 0.2113\ldots $ which  by (6) are predicted to be at the transition curve. The decay of the graph shows that they are on the ordered phase region.

\

\noindent
Figure~4. (Color online)  The graphs of $ d_H(t) $ {\it vs.} $ t $, all for $ K = 3 $  and the bias values $ p = 0.22 $ (red), $ p = 0,23 $ (green), and $ p = 0.2654\ldots $ (blue) in increasing order of their slopes. All are increasing graphs, particularly the theoretically predicted values $ p = 0.2654\ldots $ and $ K = 3 $ which by (30), should be on the transition curve. Its grow, however, is interpreted an artefact due the finite value of $ N $ according to (32).

\

\noindent
Figure~5. (Color online)  The graphs of $ d_H(t) $ {\it vs.} $ t $ for the connectivity value $ K = 3 $ and the bias values $ p = 0.3 $ (red), and $ p = 0.5 $ (green) in increasing order of their slopes. They show an initial exponential grow typical of the disordered phase, and then they begin to change to a constant value, due to the saturation effect when $ d_H (t) \sim N / 4 $.

\

\noindent
Figure~6. (Color online) The graph of $ d_H(t) $ {\it vs.} $ t $ for the controversial parameter's values $ K = 2 $, $ p = 1 / 2 $. The $ d_H(t) $ decay shows an ordered phase. The grow of the error bars is due to the decrease of $ d_H(t) $ which increases the relative error by $ 1 / \sqrt{d_H(t)} $ according to (27).

\

\noindent
Figure~7. (Color online) The graphs of $ d_H(t) $ {\it vs.} $ t $, all for the connectivity value $ K = 3 $, and the vias' values $ p = 0.2113\ldots, \, 0.22, \, 0.23, \, 0.24 $, $ 0.2654 \ldots $ beginning at the bottom with $ p = 0.2113\ldots $ and going upwards for growing values of $ p $.

\end{document}